\documentclass[a4paper,twoside]{article}

\usepackage{url}
\usepackage{epsfig}
\usepackage{subcaption}
\usepackage{calc}
\usepackage{amssymb}
\usepackage{amstext}
\usepackage{amsmath}
\usepackage{amsthm}
\usepackage{multicol}
\usepackage{multirow}
\usepackage{pslatex}
\usepackage{apalike}
\usepackage{algorithm2e}
\usepackage[bottom]{footmisc}
\usepackage{SCITEPRESS}

\begin{document}

\title{Zero-shot Evaluation of Deep Learning for Java Code Clone Detection}

\author{\authorname{Thomas S. Heinze\sup{1}}
\affiliation{\sup{1}Cooperative University Gera-Eisenach, Gera, Germany}
\email{thomas.heinze@dhge.de}}


\abstract{Deep Learning (DL) is becoming more and more widespread in clone
          detection, motivated by achieving near-perfect performance for this
          task. In particular in case of semantic code clones, which share
          only limited syntax but implement the same or similar functionality,
          Deep Learning appears to outperform conventional tools. In this
          paper, we want to investigate the generalizability of DL-based
          clone detectors for Java. We therefore replicate and evaluate the
          performance of five state-of-the-art DL-based clone detectors,
          including Transformers like CodeBERT and single-task models like
          FA-AST+GMN, in a zero-shot evaluation scenario, where we
	  train/fine-tune and evaluate on different datasets and
          functionalities. Our experiments demonstrate that the models’
          generalizability to unseen code is limited. Further analysis
          reveals that the conventional clone detector NiCad even outperforms
          the DL-based clone detectors in such a zero-shot evaluation
          scenario.}

\onecolumn \maketitle \normalsize \setcounter{footnote}{0} \vfill

\section{\uppercase{Introduction}}

\emph{Code clones}, ranging from duplicated code to semantic code clones,
i.e., syntactically different fragments of code with same functionality,
are frequently found in today's code bases and can influence aspects like
maintainability, code quality, and defect/vulnerability
proneness~\cite{RoyCK09}. While detection of identical or near-miss code
clones is considered a largely solved problem, identifying semantic clones
remains challenging. In particular, Deep Learning (DL) approaches address
this gap and show near-perfect results, as is usually demonstrated on
\emph{BigCloneBench}~\cite{WangLM0J20,LuGRHSBCDJTLZSZ21,GuoLDW0022}. As
such evaluations are often conducted within this single benchmark,
generalizability of the DL models to unseen functionalities, i.e., code
implementing behaviour absent in the models' training data, remains
disputed. We therefore want to investigage the models under
\emph{zero-shot evaluation} referring to a scenario where a DL-based
clone detector is tested or used on code it was never trained on.

Our main contributions are: We present replication and evaluation experiments
on the performance of Deep Learning clone detectors for Java, including
\emph{CodeBERT}~\cite{FengGTDFGS0LJZ20},
\emph{GraphCodeBERT}~\cite{GuoRLFT0ZDSFTDC21},
\emph{UniXcoder}~\cite{GuoLDW0022}, \emph{CodeT5}~\cite{0034WJH21}, and
\emph{FA-AST+GMN}~\cite{WangLM0J20}, as well as the conventional tools
\emph{NiCad}~\cite{RoyC08a}, \emph{NIL}~\cite{NakagawaHK21}, and
\emph{StoneDetector}~\cite{AmmeHS21}. We in particular provide comprehensive
analysis of their generalizability using four different evaluation benchmarks
from varying application scenarios and domains, besides BigCloneBench,
and demonstrate a general drop of on average approx. 41\% in the models'
F1 scores under zero-shot evaluation.

We believe our paper to be interesting to both, researchers and practitioners,
as it shows: (1) clone detectors' performance evaluations require careful
analysis and should not be solely based on BigCloneBench, (2) the
threshold value of conventional tools like NiCad is a beneficial
configuration parameter when it comes to semantic code clones, and (3) there
is no ``free lunch'', i.e., there is no clone detector which outperforms all
the others in every domain.

\section{\uppercase{Benchmarks}} \label{sec:benchmarks}

A number of datasets exists with code clone samples for the Java programming
language which can be used for benchmarking clone detectors. While
BigCloneBench is the de-facto standard~\cite{SvajlenkoR15}, its usage
for training/fine-tuning and evaluating Deep Learning approaches to code
clone detection is rather controversial~\cite{KrinkeR22,SchaferAH22}.
We therefore use a number of benchmarks beside BigCloneBench in our
evaluation experiments, covering various application contexts and domains,
including open-source production code, submissions to programming contests,
and code snippets from Q\&A fora (cf. Table~\ref{tbl:benchmarks}). To
foster further research and for replication, we provide all five benchmark
datasets online\footnote{https://doi.org/10.5281/zenodo.19581107}.

\begin{table}[t]
    \caption{Code clone detection benchmarks in this paper.}
    \label{tbl:benchmarks}
    \scriptsize \centering
    \begin{tabular}[h]{c|c|c|c}
        \multirow{2}{*}{Benchmark} &
        \#Code &
        \#Positive &
        \#Negative \\
        & Fragments & Samples & Samples \\ \hline 
        \emph{BigCloneBench} & 9,126 & 56,820 & 358,596 \\
        \emph{SemanticCloneBench} & 1,000 & 1,000 & 1,000 \\
        \emph{FEMPD} & 4,388 & 1,342 & 852 \\
        \emph{SeSaMe} & 1,217 & 66 & 546 \\
        \emph{ProjectCodeNet} & 2,919 & 1,000 & 1,000
    \end{tabular}
\end{table}

\paragraph{BigCloneBench:} Svajlenko et al. introduce the
\emph{BigCloneBench} dataset for evaluating performance, in particular
recall, of clone detectors for the Java programming language and provide 
the de-facto standard dataset in code clone research~\cite{SvajlenkoR15}.
BigCloneBench has been derived from the inter-project source code dataset
\emph{IJADataset 2.0}, comprising approx. 365 million lines of code in more
than 2.3 million Java source code files from 25,000 open-source projects.
The authors of BigCloneBench use a multi-step approach to create
the dataset which is centered around 43 functionalities, e.g., copy a file
or web download. Using heuristics, candidate methods have been mined from
the IJADataset 2.0 and manually classified according to the 43
functionalities. Code clones are generated from methods which have
been assigned the same functionality. Eventually, BigCloneBench
contains more than 8 million known code clones of different syntactical
similarity.

In recent years, various research relied on this dataset for Deep Learning
approaches to clone detection (cf.~\cite{KrinkeR22,KrinkeR25}). Instead of
the whole dataset, oftentimes, a subset of BigCloneBench is
used~\cite{WeiL17,WangLM0J20,LuGRHSBCDJTLZSZ21}. Starting with the authors
of the Deep Learning clone detector CDLH~\cite{WeiL17}, this subset is
constructed by discarding those ``... code fragments without any tagged true
and false clone pairs'', yielding approx. 9,100 Java methods. While the
positive samples of Java code clones can then be simply derived from
BigCloneBench, the construction of negative samples is though kept
quite opaque. In our experiments, we use the \emph{CodeXGLUE}
variant~\cite{LuGRHSBCDJTLZSZ21}\footnote{\url{https://github.com/microsoft/CodeXGLUE}}
of this unbalanced subset as baseline, which comprises 1,731,860
methods pairs over 9,126 Java methods from BigCloneBench, including
561,521 positive and 1,170,339 negative samples of Java code clones split
across training, validation, and evaluation datasets
(Table~\ref{tbl:benchmarks} refers to CodeXGLUE's evaluation dataset).
We note that the usage of BigCloneBench for training/fine-tuning and
evaluating Deep Learning approaches is disputed within the
literature~\cite{KrinkeR22} (cf. Sect.~\ref{sec:relatedwork}).
After all, we hope to contribute a further clarification to this dispute
with our work.

\paragraph{SemanticCloneBench:}
\emph{Stack Overflow}\footnote{\url{https://stackoverflow.com}} is a community
web platform, which allows users to ask and answer questions related to
various programming topics. Al-Omari et al.~\cite{Al-OmariRC20} used Stack
Overflow as a source for their \emph{SemanticCloneBench}
dataset\footnote{\url{https://drive.google.com/open?id=1KicfslV02p6GDPPBjZHNlmiXk-9IoGWl}}
of (semantic) code clones based on the idea that code snippets in correct
answers to the same question are functionally similar and thus constitute
a code clone. They apply additional steps, e.g., filtering out syntactical
clones and manually validating the identified clones by two judges, which
results in overall 4,000 clone pairs, including 1,000 code clones for the
Java programming language. While the dataset itself only contains positive
samples of code clones, Arshad et al. later proposed a simple approach for
generating negative samples in~\cite{ArshadAS22}: Their idea is to combine
each first element of a clone pair in the dataset's first half with a first
element of the clone pairs in the dataset's second half and to do so
similarly for the clone pairs' second elements. We use the same approach
and are able to construct a balanced dataset containing 1,000 positive and
1,000 negative samples for Java code clones. 

\paragraph{FEMPD:} The benchmark
\emph{FEMPD}~\cite{Higo24}\footnote{\url{https://github.com/YoshikiHigo/FEMPDataset}}
defines a dataset of in particular semantic code clones. The dataset has
been generated on the inter-project source code repository IJADataset,
following a rigorous approach using various steps, including grouping Java
methods according to their static signatures, generating and running test
cases for identifying functionally equivalent methods, and manually
validating the thus determined semantic clone pairs. This approach results
in an unbalanced dataset which contains 1,342 positive samples and 852
negative samples for Java code clones. Note that FEMPD originates
from the IJADataset, much like BigCloneBench, though is
based on a more strict and rigorous definition of code clone, i.e.,
functionally equivalent code.

\paragraph{SeSaMe:} Another dataset containing examples of real-world
Java code clones is
\emph{SeSaMe}~\cite{KampKP19}\footnote{\url{https://github.com/FAU-Inf2/sesame}}.
Its authors focus on semantically similar code fragments from real-world
production code and therefore mined large open-source Java projects,
including, e.g., Eclipse's Java Development Tools (JDT), Google's Guava
library, and the Open Java Development Toolkit. As a starting point,
they consider the API documentations and analyze methods' documentation
comments for textual similarity. In a subsequent step, they manually
assess the resulting method pairs according to three similarity dimensions,
i.e., methods' goals, operations and effects. In contrast to the other
datasets included in our experiments, SeSaMe does not define a binary
classification of method pairs into positive and negative samples of code
clones, but rather contains 857 method pairs conjoined with respective
manual judgements of their similarity scores. Accordingly, we derive
an unbalanced dataset set of 66 positive samples and 546 negative samples
for Java code clones by only considering those method pairs which feature
a majority rate as similar and a majority rate as dissimilar in all three
similarity dimensions, respectively.

\paragraph{ProjectCodeNet:} The archives of online programming contests
like \emph{Google Code Jam}\footnote{\url{https://zibada.guru/gcj/}} or
\emph{AtCoder}\footnote{\url{https://atcoder.jp}} provide a rich source
of semantic code clones and have therefore been utilized for the definition
of clone datasets and benchmarks. We include
\emph{ProjectCodeNet}~\cite{Puri0JZDZD0CDTB21}\footnote{\url{https://github.com/IBM/Project_CodeNet}}
in our experiments as a representative example due to its large size and
inclusion of Java code. The dataset originates from the \emph{AIZU} and
\emph{AtCoder} programming contests and contains in its Java benchmark
subset 750,000 submissions to 250 programming tasks. Prior cleansing
filters out identical problems and near-duplicate submissions. Two accepted
submissions to the same task then constitute a positive sample of code
clones whereas two accepted submissions to different tasks constitute a
negative sample. As each submission consists of a single Java source file
with potentially more than one Java method, we additionally filter for
submissions comprising a single method. In this way, we were able to
construct a balanced dataset with 1,000 positive and 1,000 negative
samples of Java code clones.

\section{\uppercase{Clone Detectors}} \label{sec:tools}

Clone detection is a well-studied research subject and numerous clone
detectors have been proposed in the literature~\cite{RoyCK09}. While most
clone detectors are conventional, with the advent of Deep Learning, more
and more tools employ this approach for finding code clones. In this
section, we will shortly introduce the selected conventional and
Deep Learning-based clone detectors used in our experiments.

\subsection{DL-based Clone Detectors}

\begin{table}[t]
    \caption{Deep Learning-based clone detectors in this paper.}
    \label{tbl:models}
    \scriptsize \centering
    \begin{tabular}[h]{c|c|c}
        Model & Model Type & \#Parameters \\ \hline
        \multirow{2}{*}{\emph{CodeBERT}} &
        pre-trained, masked language &
        \multirow{2}{*}{125 million} \\
        & model (encoder-only) & \\
        \multirow{2}{*}{\emph{GraphCodeBERT}} &
        pre-trained, masked language &
        \multirow{2}{*}{125 million} \\
        & model (encoder-only) & \\
        \multirow{2}{*}{\emph{UniXcoder}} &
        pre-trained, unified multi- &
        \multirow{2}{*}{125 million} \\
        & mode transformer model & \\
        \emph{CodeT5-base} & pre-trained, seq-to-seq model & 220 million \\
        \emph{FA-AST+GMN} & graph-based neural network & n/a
    \end{tabular}
\end{table}

The Transformer architecture and pre-trained general-purpose code models have
been shown to achieve promising results in various programming language
tasks, including clone detection. We select four different Transformer models
to provide for a comprehensive picture of their capabilities (cf.
Table~\ref{tbl:models}).

\emph{CodeBERT}~\cite{FengGTDFGS0LJZ20},
\emph{GraphCodeBERT}~\cite{GuoRLFT0ZDSFTDC21}, and
\emph{UniXcoder}~\cite{GuoLDW0022}\footnote{\url{https://github.com/microsoft/CodeBERT}}
are pre-trained models for code, i.e., they have been pre-trained on large
code corpora and can be fine-tuned for a specific downstream task like clone
detection. While CodeBERT and GraphCodeBERT are masked language
models, i.e., are pre-trained for predicting masked code from surrounding
context, \emph{CodeT5}~\cite{0034WJH21}\footnote{\url{https://github.com/salesforce/CodeT5}}
is another pre-trained Transformer model, which though is pre-trained as
sequence-to-sequence model for auto-regressively translating an input (code)
sequence into an output (code) sequence. UniXcoder provides a uniform
multi-mode model, which has been pre-trained for multiple training
objectives.

In addition, we consider with
\emph{FA-AST+GMN}~\cite{WangLM0J20}\footnote{\url{https://github.com/jacobwwh/graphmatch_clone}}
a representative of a single-task neural network model, which falls into the
same family like CDLH~\cite{WeiL17}, ASTNN~\cite{ZhangWZ0WL19},
in contrast to above's pre-trained general-purpose language models to
complete the picture. FA-AST+GMN finds code clones by representing
two code fragments by data flow‑augmented abstract syntax trees and then
using graph‑matching neural networks to embed and match the two graphs
based on their cosine similarity.

Note that the five models are usually trained/fine-tuned and evaluated on
subsets of BigCloneBench and then regularly achieve F1 scores of
$\geq 0.94$~\cite{WangLM0J20,LuGRHSBCDJTLZSZ21,GuoLDW0022}.

\subsection{Conventional Tools}

\begin{table}[t]
    \caption{Conventional code clone detectors in this paper conjoined with
             their used configuration parameters.}
    \label{tbl:tools}
    \scriptsize \centering
    \begin{tabular}[h]{l|l|l}
        \multicolumn{1}{c|}{\emph{NiCad v7.0.1}} &
        \multicolumn{1}{c|}{\emph{NIL v2.0.0}} &
        \multicolumn{1}{c}{\emph{StoneDetector}} \\ \hline
        70\% sim. threshold &
        10\% filtr. threshold, &
        70\% sim. threshold \\
        ($\tau$=0.3), blind renaming, &
        70\% ver. threshold &
        ($\tau$=0.3), LCS metric, \\
        literal abstraction &
        5-grams &
        8-byte hashing
    \end{tabular}
\end{table}

For sake of comparison, we include three conventional code clone detectors
for Java in our experiments (cf. Table~\ref{tbl:tools}):
\emph{NiCad}~\cite{RoyC08a} is a hybrid clone detector employing
normalization techniques ahead of analyzing code similarity based on the
normalized code fragments' longest common subsequence (LCS). NiCad is
in particular good at finding near-miss code clones at very high precision.
We use the tool's most recent free and open version 7.0.1
as available online\footnote{\url{https://github.com/CordyJ/Open-NiCad}}.
More recent tools like
\emph{NIL}~\cite{NakagawaHK21}\footnote{\url{https://github.com/kusumotolab/NIL}}
focus on large-gap code clones with many consecutive code edits or
modifications scattered around the code. NIL represents code by
N-grams derived from normalized token sequences and thereon measures
similarity again using LCS.
\emph{StoneDetector}~\cite{AmmeHS21,HeinzeSA26}\footnote{\url{https://github.com/StoneDetector/StoneDetector}}
is another recent clone detector for Java which has been shown to in
particular excel at finding code clones with larger syntactical
variance~\cite{HeinzeSA26} and for that purpose employs string metrics like
LCS on fingerprints as derived from code fragments' dominator trees.

\section{\uppercase{Evaluation}}

In our experiments, we want to investigate the generalizability of DL-based
clone detectors trained on BigCloneBench. We therefore first conduct
a replication experiment, where the five DL models introduced in
Sect.~\ref{sec:tools} are trained/fine-tuned and evaluated on the
CodeXGLUE subset of BigCloneBench, which provides us with a
baseline (cf. Table~\ref{tbl:replication}). We then use the thus trained
models and evaluate them in a zero-shot evaluation approach for the four
other benchmarks introduced in Sect.~\ref{sec:benchmarks} (cf.
Table~\ref{tbl:evaluation}). We in addition include the three conventional
tools for comparison.

All experiments were conducted on a Ubuntu 24.04 LTS system running in a
virtual machine with assigned 8 CPU cores 2.3 GHz, 48 GB RAM, and NVIDIA RTX
6000 Ada GPU (CUDA v13.0).

\subsection{Evaluation Metrics}

Clone detection can be seen as a binary classification problem, where a pair
of code fragments is assgined one of two categories: code clone (positive
assignment) or non-clone (negative assignment). Consequently, for a certain
clone detector and code clone dataset, we can differentiate between the clone
detector's correct positive assignments, i.e., \emph{true positives}, correct
negative assignments, i.e., \emph{true negatives}, incorrect positive
assignments, i.e., \emph{false positives}, and incorrect negative assignments,
i.e., \emph{false negatives}. \emph{Recall} and \emph{precision} are then
standard evaluation metrics for assessing the probability of detecting a
true clone and the propability of a correct positive classification using
the clone detector, respectively:

\begin{eqnarray*}
    \mathit{Recall} & = & \frac{\#\mbox{True Positives}}
                        {\#\mbox{True Positives}+\#\mbox{False Negatives}} \\
    \mathit{Precision} & = & \frac{\#\mbox{True Positives}}
                        {\#\mbox{True Positives}+\#\mbox{False Positives}} \\
\end{eqnarray*}

Note that a trivial clone detector, which assigns each pair of code fragments
as code clone, can achieve perfect recall and -- vice versa -- a clone
detector, which assigns each pair as non-clone, can achieve perfect
precision. Thus, assessing a clone detectors' performance requires to analyze
both metrics. In addition, the \emph{fall-out} or false-positive rate may be
used as measure for the probability of false alarms:

\begin{eqnarray*}
    \mathit{Fall-out} & = & \frac{\#\mbox{False Positives}}
                          {\#\mbox{False Positives}+\#\mbox{True Negatives}}
\end{eqnarray*}

Averaging recall and precision into a single evaluation metric can be done
using their harmonic mean, i.e., \emph{F1 score}, as follows:

\begin{eqnarray*}
    \mathit{F1\ Score} & = & 2\times \frac{\mbox{Precision}\times\mbox{Recall}}
                            {\mbox{Precision}+\mbox{Recall}}
\end{eqnarray*}

Note though that the F1 score assumes equal importance of recall and precision
and may be less informative when compared to using the other two metrics.
Furthermore, certain clone detectors support a threshold value which allows
to define the clone detector's permissiveness of false positives. In such
cases, the performance of the clone detector can be illustrated in terms of
its \emph{receiver operating characteristic (ROC) curve}. The ROC curve
plots recall and fall-out, i.e., true-positive and false-positive rate,
respectively, at varying threshold values. In this plot, a random
classification, which assigns a code clone by flipping a coin, results in a
point on the diagonal line, i.e., true positive rate equals false positive
rate. The better a clone detector, the farer is the clone detector's
characteristic function from this diagonal line. As the ROC curve allows
for evaluating a clone detector for different threshold values, it is
apparently more informative when compared to using precision, recall,
and F1 score for a single configuration alone.

\subsection{Experimental Results}

As a first step, we replicate the CodeXGLUE benchmark (cf.
Sect.~\ref{sec:benchmarks}),  training/fine-tuning and evaluating DL-based
clone detectors and just evaluating the conventional clone detectors on the
same subset of BigCloneBench. As shown in Table~\ref{tbl:replication},
we can reproduce the results as reported in the literature, i.e., all five
DL models achieve precision, recall, and F1 scores above 0.9. As expected,
the conventional clone detectors in comparison only detect a small fraction
of the code clones, resulting in a very low recall, while achieving similar
precision scores above 0.9. In the table, we also provide the tools' runtimes
and the DL models' used GPU memory. As expected, CodeT5 is the largest
model and fine-tuning the pre-trained models takes considerably less time
than full-training of FA-AST+GMN. Note that evaluating the samples of
CodeXGLUE's evaluation subset takes at least one hour in case of the
DL models while lasting approx. one minute in case of the conventional tools.

\begin{table}[t]
    \caption{Experimental results for \emph{BigCloneBench} (R -- recall,
             P -- precision, GPU memory and fine-tuning/evaluation time is
             given if applicable and available).}
    \label{tbl:replication}
    \scriptsize \centering
    \begin{tabular}[h]{c|c|c|c|c|c}
        Clone Detector & R & P & F1 & Runtime & GPU \\
        \hline
        \multirow{2}{*}{\emph{CodeBERT}} &
        \multirow{2}{*}{0.96} &
        \multirow{2}{*}{0.92} &
        \multirow{2}{*}{0.94} &
        77 min/ &
        \multirow{2}{*}{10,549 MiB} \\
        & & & & 64 min & \\
        \emph{GraphCode} &
	\multirow{2}{*}{0.95} &
        \multirow{2}{*}{0.94} &
        \multirow{2}{*}{0.95} &
	634 min/ &
        \multirow{2}{*}{15,941 MiB} \\
        BERT & & & & 101 min & \\
	\multirow{2}{*}{\emph{UniXcoder}} &
	\multirow{2}{*}{0.95} &
        \multirow{2}{*}{0.93} & 
        \multirow{2}{*}{0.94} &
        444 min/ &
        \multirow{2}{*}{12,833 MiB} \\
        & & & & 73 min & \\
        \emph{CodeT5} & 0.94 & 0.96 & 0.95 & 1,368 min & 31,973 MiB \\
        \emph{FA-AST+GMN} & 0.94 & 0.93 & 0.93 & 2,662 min & 5,377 MiB \\
        \hline
        \emph{NiCad v7.0.1} & 0.01 & 0.92 & 0.01 & 1 min & - \\
        \emph{NIL} & 0.01 & 0.91 & 0.02 & 1 min & - \\
        \emph{StoneDetector} & 0.01 & 0.90 & 0.02 & 1 min & -
    \end{tabular}
\end{table}
\bigskip

Training or fine-tuning the Deep Learning models on the BigCloneBench
subset and evaluating them on one of the other benchmarks, i.e.,
FEMPD, SeSaMe, SemanticCloneBench, or
ProjectCodeNet yet paints a different picture. Like mentioned above,
we want to investigate on the models' generalizability using this zero-shot
evaluation scenario. As can be seen in Table~\ref{tbl:evaluation},
performance deteriorates for all five DL models on all four
benchmarks with F1 scores dropping on average by approx. 41\%. For instance,
while recall remains on the same level or
drops at most to 0.74 for CodeBERT, its precision shrinks on average
to 0.46. Note that a precision score 0.5 equals flipping a coin for deciding
whether an identified code clone is indeed a code clone or not. The other
Transformer models achieve better precision but lower recall compared to
CodeBERT and there is apparently not a single model which outperforms
the other ones on all benchmarks. Furthermore, we observe that the
Transformer models' performance in particular degrades for benchmark
SeSaMe, which is striking considering the benchmark's origin in
open-source production projects and its ground-truth quality
(Sect.~\ref{sec:benchmarks}). Also remarkable, the single-task model
FA-AST+GMN achieves the best precision among the five DL clone
detectors over the four benchmarks.

\begin{table}[t]
    \caption{Experimental results for benchmarks \emph{SeSaMe},
             \emph{SemanticCloneBench}, \emph{FEMPD}, and
             \emph{ProjectCodeNet} (R -- recall, P -- precision).}
    \label{tbl:evaluation}
    \scriptsize \centering
    \begin{tabular}[h]{c|c|c|c|c|c|c}
        \multirow{2}{*}{Clone Detector} &
        \multicolumn{3}{c|}{\emph{SemanticCloneBench}} &
        \multicolumn{3}{c}{\emph{SeSaMe}} \\
        & R & P & F1 & R & P & F1 \\
        \hline
        \emph{CodeBERT} & \textbf{0.74} & 0.54 & 0.62 & \textbf{0.94}& 0.15 & 0.26 \\
        \emph{GraphCode} &
        \multirow{2}{*}{0.45} & 
        \multirow{2}{*}{0.72} &
        \multirow{2}{*}{0.56} &
        \multirow{2}{*}{0.39} &
        \multirow{2}{*}{0.33} &
        \multirow{2}{*}{0.36} \\
        \emph{BERT} & & & & & & \\
        \emph{UniXcoder} & 0.53 & 0.78 & \textbf{0.63} & 0.55 & 0.55 & 0.55 \\
        \emph{CodeT5} & 0.28 & 0.78 & 0.41 & 0.11 & 0.29 & 0.16 \\
        \emph{FA-AST+GMN} & 0.38 & 0.80 & 0.52 & 0.53 & 0.61 & 0.57 \\
        \hline
        \emph{NiCad v7.0.1} & 0.02 & \textbf{1.0} & 0.04 & 0.45 & \textbf{1.0} & 0.63 \\ 
        \emph{NIL} & 0.14 & 0.99 & 0.25 & 0.53 & \textbf{1.0} & \textbf{0.69} \\
        \emph{StoneDetector} & 0.05 & \textbf{1.0} & 0.09 & 0.30 & \textbf{1.0} & 0.47
    \end{tabular}
    \smallskip \\ 
    \begin{tabular}[h]{c|c|c|c|c|c|c}
        \multirow{2}{*}{Clone Detector} &
        \multicolumn{3}{c|}{\emph{FEMPD}} &
        \multicolumn{3}{c}{\emph{ProjectCodeNet}} \\
        & R & P & F1 & R & P & F1 \\
        \hline
        \emph{CodeBERT} & \textbf{0.97} & 0.62 & \textbf{0.76} & \textbf{0.81} & 0.53 & \textbf{0.64} \\
        \emph{GraphCode} &
        \multirow{2}{*}{0.65} &
        \multirow{2}{*}{0.64} &
        \multirow{2}{*}{0.65} &
        \multirow{2}{*}{0.53} &
        \multirow{2}{*}{0.61} &
        \multirow{2}{*}{0.57} \\
        \emph{BERT} & & & & & & \\
        \emph{UniXcoder} & 0.68 & 0.67 & 0.67 & 0.78 & 0.51 & 0.62 \\
        \emph{CodeT5} & 0.40 & 0.67 & 0.50 & 0.73 & 0.56 & 0.63 \\
        \emph{FA-AST+GMN} & 0.54 & \textbf{0.70} & 0.61 & 0.17 & 0.75 & 0.28 \\
        \hline
        \emph{NiCad v7.0.1} & 0.18 & 0.62 & 0.28 & 0.02 & \textbf{1.0} & 0.04 \\
        \emph{NIL} & 0.34 & 0.66 & 0.45 & 0.19 & 0.87 & 0.31 \\
        \emph{StoneDetector} & 0.41 & \textbf{0.70} & 0.52 & 0.10 & 0.98 & 0.18
    \end{tabular}
\end{table}

In contrast, we observe in general a slightly better recall for the three
conventional clone detectors NiCad, NIL, and
StoneDetector, as well as very good or acceptable precision scores
with exclusion of benchmark FEMPD. Obviously, conventional tools do
not suffer from the same generalizability problem as the Deep Learning-based
tools. Interestingly, they achieve their best overall performance on
SeSaMe, where NiCad and NIL with F1 scores of 0.63,
0.69, respectively, even outperform the DL-based models.

\begin{figure}[t]
    \centerline{\includegraphics[width=\columnwidth]{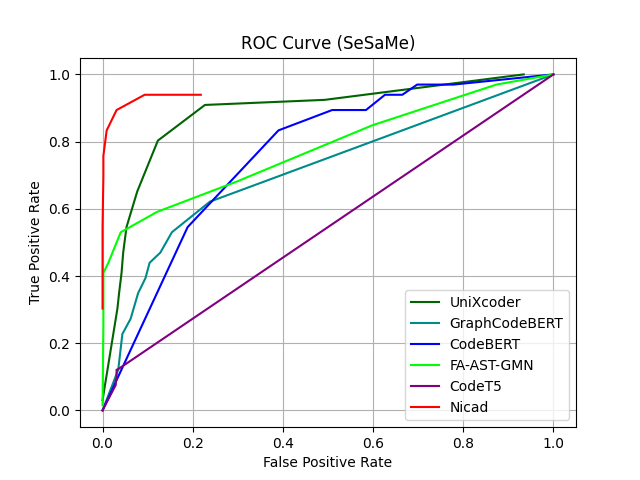}}
    \caption{ROC curve for DL-based clone detectors and conventional clone
             detector \emph{NiCad} on benchmark \emph{SeSaMe}.}
    \label{fig:sesame}
\end{figure}

\begin{figure}[t]
    \centerline{\includegraphics[width=\columnwidth]{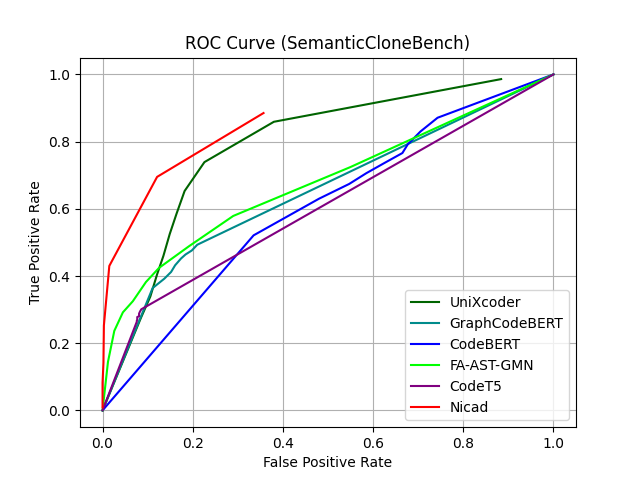}}
    \caption{ROC curve for DL-based clone detectors and conventional
             tool \emph{NiCad} on benchmark \emph{SemanticCloneBench}.}
    \label{fig:semanticclonebench}
\end{figure}

\begin{figure}[t]
    \centerline{\includegraphics[width=\columnwidth]{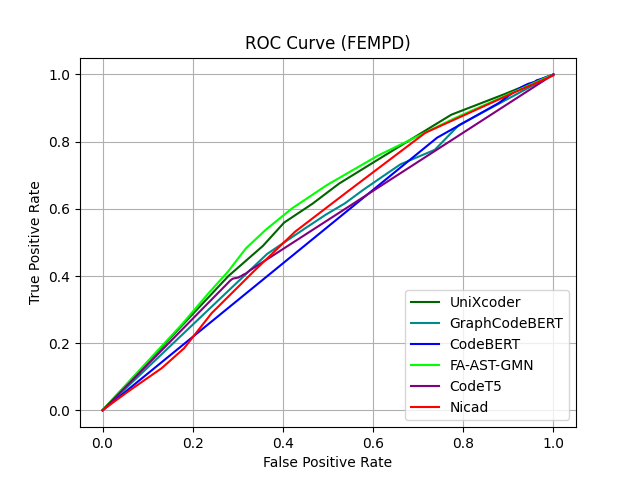}}
    \caption{ROC curve for DL-based clone detectors and conventional
             tool \emph{NiCad} on benchmark \emph{FEMPD}.}
    \label{fig:fempd}
\end{figure}

\begin{figure}[t]
    \centerline{\includegraphics[width=\columnwidth]{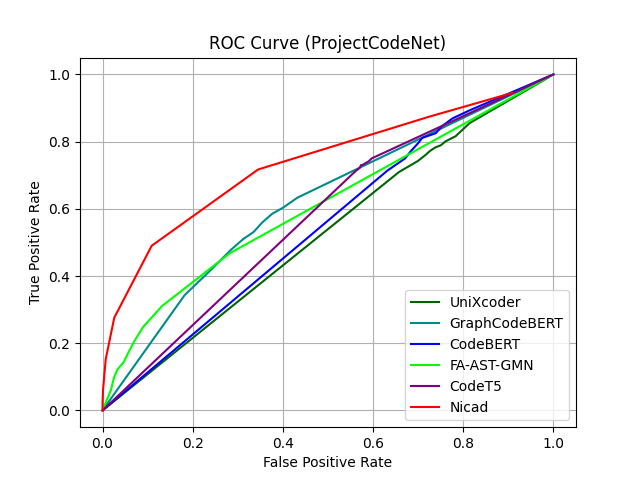}}
    \caption{ROC curve for DL-based clone detectors and conventional
             tool \emph{NiCad} on benchmark \emph{Project Codenet}.}
    \label{fig:codenet}
\end{figure}

In the beginning of the section, we argued that just using precision, recall,
and F1 score does not provide for a sufficient discussion of clone detectors'
performance in cases where they support a threshold value and that the ROC
curve may then be used. We extend our experiments to track the five DL models
recall and fall-out for different threshold values (ranging from 0.0 to 1.0
with the exception of FA-AST+GMN, where it ranges from -1.0 to 1.0).
We additionally provide the ROC curve of the conventional tool NiCad
for comparison ($\tau\in [0.0,1.0]$, c.f. Sect~\ref{sec:tools}). The
resulting ROC curves are given in Fig.~\ref{fig:sesame} to
Fig.~\ref{fig:codenet}. Most apparent, FEMPD yields similar
curves for all tools, indicating poor performance, which is attributed
to its focus on functionally equivalent code, which seems harder to identify
(cf. Sect.~\ref{sec:benchmarks}). Second, each benchmark has its own
characteristic curves and we again do not find one DL model which outperforms
the others for all benchmarks, while UniXcoder and FA-AST+GMN
show better curves on average (the closer to the upper left the better).
Eventually, and maybe unexpectedly, NiCad shows superior sensitivity
and specificity in all benchmarks beside FEMPD.

\section{\uppercase{Related Work}} \label{sec:relatedwork}

While BigCloneBench has been widely used for training and evaluating
Deep Learning clone detectors for Java, BigCloneBench's suitability
for this purpose became disputed in recent years. Krinke et al. focus on the
general benchmark's ground truth quality and its widespread usage as training
dataset in~\cite{KrinkeR22,KrinkeR25}. They specifically identify the
benchmark's overlapping functionalities, invalid positive samples of code
clones, bias and imbalance with respect to functionalities and semantic code
clones as issues impairing its usage. Note that Krinke et al. do not provide
experimental analysis of these issues besides a manual investigation of a
random sample of BigCloneBench' code clones. We hope to provide, with
our work, comprehensive experimental evidence to enrich this ongoing
discussion.

\begin{table}[t]
    \caption{Related work on Java code clone detection with
             \emph{CodeBERT} (R -- Recall, P -- Precision, BCB --
             \emph{BigCloneBench}, SCB -- \emph{SemanticCloneBench},
             superscripts \textsuperscript{*,\dag,\P,\S} indicate
             different subsets of \emph{BigCloneBench}).}
    \scriptsize \centering
    \begin{tabular}[h]{l|c|c|c|c|c}
        & Train. & Eval. & R & P & F1 \\ \hline
        \cite{LuGRHSBCDJTLZSZ21} & \emph{BCB} & \emph{BCB}
        & n/a & n/a & 0.94 \\ \hline
        \cite{SonnekalbGBM22} & \emph{BCB} & \emph{BCB}\textsuperscript{*}
        & 0.52 & 0.98 & 0.68 \\
        \cite{SonnekalbGBM22} & \emph{BCB} & \emph{BCB}\textsuperscript{**}
        & 0.33 & 0.98 & 0.49 \\
        \cite{PinkuMR24} & \emph{BCB}\textsuperscript{\S} & \emph{SCB}
        & 0.47 & 0.70 & 0.56 \\
        \cite{KitsiosSBB25} & \emph{BCB}\textsuperscript{$\dagger$}
        & \emph{BCB}\textsuperscript{$\ddagger$} & 0.84 & 0.91 & 0.86 \\
        \cite{KitsiosSBB25} & \emph{BCB}\textsuperscript{$\dagger$}
        & \emph{SCB} & 0.50 & 0.96 & 0.66 \\
        \cite{ArshadAS22} & \emph{BCB}\textsuperscript{\P} & \emph{SCB}
        & 0.73 & 0.53 & 0.61 \\
        \cite{ArshadAS22} & \emph{BCB}\textsuperscript{\P} & \emph{Android}
        & 0.64 & 0.87 & 0.74 \\ \hline
        our paper & \emph{BCB} & \emph{SCB}
        & 0.74 & 0.54 & 0.62
    \end{tabular}
\end{table}

Some research experimentally analyze the suitability of BigCloneBench
and in particular subsets thereof for Deep Learning of clone detectors:
In~\cite{SonnekalbGBM22}, the authors evaluate a CodeBERT model
fine-tuned for clone detection using the
CodeXGLUE subset. However, they use their own evaluation datasets,
which are derived from BigCloneBench, making sure to rule out code
duplicates or the same functionalities. As a result, they report a drop of
CodeBERT's recall from 0.96 to 0.52 and 0.33, respectively, while
precision even improves. Note that in our experiments, we rather observe a
degradation of CodeBERT's precision and not recall when using unseen
data for evaluation. In the same vein, Sch\"afer et al. investigate the
impact of using a more rigorous segregation of training and evaluation data
for FA-AST+GMN on BigCloneBench in~\cite{SchaferAH22}. Using
samples of different functionalities in training and evaluation data results
in a drop of both, recall and precision, and deteriorates the model's F1
score from 0.95 to 0.72. Note that they apply FA-AST+GMN on a
register-based intermediate representation of Java Bytecode instead of
Java's source code and also use the whole BigCloneBench and not a
subset thereof. An analysis of the DL-based clone detectors
ASTNN~\cite{ZhangWZ0WL19} and TBCCD for the C programming
language on the OJClone benchmark revealed similar
effects~\cite{LiuLLWZ21}. Its authors also investigate possible mitigations,
e.g., increasing training data diversity, addressing the out-of-vocabulary
problem, and integrating a human-in-the-loop mechanism.

In~\cite{ArshadAS22}, the authors use a CodeBERT model fine-tuned on
BigCloneBench for zero-shot evaluation, much in the same way as we do,
and report on observed significant drops of recall and precision by
15\%-44\%, a finding similar to our results. They however, in contrast to our
research, only consider CodeBERT and two evaluation datasets, i.e.,
SemanticCloneBench and an Android benchmark. They though demonstrate
that additionally fine-tuning CodeBERT on the evaluation datasets
helps to restore much of the model's prior performance. Also similarly to our
approach, Pinku et al. investigate the usage of Deep Learning to code clone
detection~\cite{PinkuMR24}. They again only consider two of the models
included in our experiments, i.e. CodeBERT and FA-AST+GMN.
They do not include conventional clone detectors for comparison and do not
examine varying threshold values. They though address with
ASTNN~\cite{ZhangWZ0WL19} another graph-based model, and with
GPTCloneBench~\cite{AlamRARRS23} another benchmark besides
SemanticCloneBench, as well as cross-language approaches. Overall,
they report similar results for training on BigCloneBench and
zero-shot evaluation on SemanticCloneBench and note a deterioration
in the F1 score (0.68 and 0.56 for FA-AST+GMN and
CodeBERT, respectively). Interestingly, they observe higher recall
and lower precision for FA-AST+GMN and higher precision and lower
recall for CodeBERT than we do.

Kitsios et al. as well look into the problem of unseen functionalities for
code clone detection with models CodeBERT, ASTNN, and
CodeGrid in~\cite{KitsiosSBB25}. They also train the models on
BigCloneBench and evaluate them on a functionality-distinct subset of
BigCloneBench and on SemanticCloneBench. A deterioration in
the models F1 score is observed, while in particular CodeBERT's recall
and ASTNN's precision is impaired. The authors also consider large
language models, i.e., GPT-4o, Llama 3.3 and DeepSeek,
with an in general worse performance when compared to CodeBERT or
ASTNN and report on contrastive learning for partially mitigating the
problem of unseen functionalities.


In~\cite{AlamRARRS25}, the authors present their findings when comparing
performance of conventional clone detectors and two Deep Learning models,
i.e., CodeBERT and ASTNN on benchmarks BigCloneBench,
GPTCloneBench, and SemanticCloneBench. Similar to us, they
observe higher recall but a much degraded precision (0.51-0.54) of Deep
learning models in comparison with conventional clone detectors like
StoneDetector~\cite{HeinzeSA26}. However, we again include a larger
number of Deep Learning models and benchmarks in our experiments and provide
further elaborations on this insight by considering varying threshold values
and the ROC metric, which is not included in~\cite{AlamRARRS25}. Like us,
they also note the better execution times and scalability of conventional
clone detectors.

\section{\uppercase{Conclusion}}

In this paper, we present our replication and evaluation experiments on
generalizability of Deep Learning approaches (DL) to Java code clone
detection. In the experiments, we analyze the detection performance of five
state-of-the-art DL models, i.e., CodeBERT, GraphCodeBERT, UniXcoder, CodeT5,
and FA-AST+GMN as trained/fine-tuned on BigCloneBench under a zero-shot 
evaluation scenario using the four benchmarks FEMPD, SemanticCloneBench,
SeSaMe, and ProjectCodeNet. We also provide an in-depth analysis of the
models' performance in comparison with the conventional tools NiCad, NIL,
and StoneDetector. Our experiments demonstrate a significant drop of the DL
models performance under zero-shot evaluation (approx. 41\% in their
F1 scores), that clone detectors' performance is coupled to the
characteristics of the used evaluation benchmark, and that the conventional
clone detector NiCad in general outperforms the DL models under zero-shot
evaluation. With our work, we hope to contribute insights for further
research and provide all datasets used in our experiments
online\footnote{\url{https://doi.org/10.5281/zenodo.19581107}}.

In future work, we want to extend our experiments with respect to more
Deep Learning models and benchmarks for Java code clone detection, e.g.,
ASTNN~\cite{ZhangWZ0WL19}, CDLH~\cite{WeiL17}, and
GPTCloneBench~\cite{AlamRARRS23}. We also want to integrate
our experiments into the \emph{CloReCo} platform~\cite{BurockAHO25}, in
order to facilitate reproducibility of clone detector performance
analysis. Eventually, we are interested in analyzing ways to improve
the performance of DL-based clone detectors on unseen code, e.g., using
techniques of domain adaptation~\cite{GrunerSHB23}.

\section*{\uppercase{Acknowledgements}}

The author would like to thank Daniel Bari\'e for providing the GPU resources
for the experiments.

\bibliographystyle{apalike}
\bibliography{references}

\end{document}